\documentclass[11pt]{article}

\usepackage[utf8]{inputenc}
\usepackage[vmargin=1.25in]{geometry}
\usepackage{amsmath,amssymb}
\usepackage{mathrsfs}
\usepackage[square]{natbib}
\usepackage{graphicx}
\usepackage[colorlinks,citecolor=green,linkcolor=blue,bookmarks=false,hypertexnames=true]{hyperref}
\usepackage{enumitem}
\def\x{\mathbf x}
\def\z{\hat{\mathbf z}}
\def\u{\mathbf u}
\def\b{\mathbf b}
\def\f{\mathbf f}
\def\J{J}
\def\R{\mathbb R}

\usepackage{titling}
\usepackage{array}
\preauthor{\begin{center}
    \large \lineskip .75em%
        \begin{tabular}[t]{>{\centering\arraybackslash}p{.45\textwidth}}}
        \postauthor{\end{tabular}\par\end{center}}
\makeatletter
\renewcommand\and{%                  % \begin{tabular}
  \end{tabular}%
  \hfill
  \begin{tabular}[t]{>{\centering\arraybackslash}p{.45\textwidth}}}%   % \end{tabular}
  \makeatother

\begin{document}

\title{Do eddies connect the tropical Atlantic Ocean and the
Gulf of Mexico?}

\author{F.\ Andrade-Canto\\ Departamento de Observaci\'on y Estudio
de la Tierra, la Atm\'osfera y el Oc\'eano\\ El Colegio de la
Frontera Sur\\ Chetumal, Quintana Roo, Mexico\\
fernando.andrade@ecosur.mx\and F.J.\ Beron-Vera\\ Department of
Atmospheric Sciences\\ Rosenstiel School of Marine \& Atmospheric
Science\\ University of Miami\\ Miami, Florida, USA\\ fberon@miami.edu}
\date{Started: February 3, 2022. This version: \today.\vspace{-0.25in}}
\maketitle

\section*{Abstract} 

Consistent with satellite-tracked trajectories of drogued drifters,
but at odds with Eulerian assessment of satellite-altimetry
measurements of sea-surface height, we show that North Brazil
Currents Rings (NBCRs) are incapable of bypassing the Lesser Antilles
as structures that coherently transport material. The nature of the
inability of the de-facto oceanographic Eulerian, streamline-based
eddy detection technique to produce a correct assessment is rooted
in its lack of objectivity.  We arrive at this conclusion by applying
\emph{geodesic eddy detection} on the altimetric dataset over nearly
its entire extent.  While we detect northwestward translating NBCRs
that can be classified as coherent Lagrangian eddies, they typically
experience strong filamentation and complete loss of coherence prior
to reaching the Lesser Antilles.   Moreover, the filamented material
hardly penetrates into the Caribbean Sea, let alone the Gulf of
Mexico, and not without substantively mixing with the ambient fluid
east of the archipelago.

\section*{Plain Language Summary}

Geodesic eddy detection is a Lagrangian (fluid-parcel-following)
method that objectively, i.e., in an observer-independent manner,
reveals eddies (fluid regions spinning about a common axis) with
material (fluid) boundaries that defy the exponential stretching
that arbitrary material loops typically experience in turbulent
flow.  Eddies instantaneously revealed as regions enclosed by
streamlines in general do not have such a property: the topology
of the streamlines depends on the observer viewpoint.  Thus
instantaneously closed streamlines cannot be guaranteed to hold and
carry within material.  Here we use geodesic eddy detection to
demonstrate that North Brazil Current rings detected from their
Eulerian footprints in the altimetric sea-surface height field do
not connect the tropical Atlantic Ocean and the Gulf of Mexico.

\section*{Key Points}

\begin{itemize}
  \item The topology of SSH streamlines is observer dependent
	and hence are eddies from the AVISO+ META database
  \item Geodesic eddy detection objectively reveals eddies with
	stretching resisting material boundaries that also minimize
	diffusion
  \item North Brazil Current Ring contents hardly penetrate into
  the Caribbean without mixing with the ambient fluid east of the
  Lesser Antilles
\end{itemize}

\section{Introduction}

\citet{Huang-etal-21} have recently argued that eddies, more
specifically mesoscale anticyclonic North Brazil Current Rings
(NBCRs), connect the tropical Atlantic Ocean and the Gulf of Mexico.
NBCRs are offsprings of the North Brazil Current retroflection
\citep{Johns-etal-90, Didden-Schott-93, Goni-Johns-01, Wilson-etal-02,
Johns-etal-03, Goni-Johns-03, Garraffo-etal-03}. As mesoscale oceanic
eddies, they have potential long-range ability of dragging along
various types of tracers (e.g., nutrients, salinity, larvae)
\citep{Robinson-83}, and, in particular, \emph{Sargassum} rafts
\citep{Beron-Miron-20}.  The argument of \citet{Huang-etal-21} is
based on the tracking of Eulerian \emph{footprints} of mesoscale
vortices on satellite-altimetry gridded maps of sea-surface height
(SSH) anomaly \citep{Schlax-Chelton-16} and absolute dynamic topography
\citep{Rio-etal-11}.  Here we refute this claim and support our
rebuttal on observations by \citet{Fratantoni-Richardson-04} of
satellite-tracked surface drifters and submerged floats, which show
that they hardly traverse the Lesser Antilles.  The main issue with
the flawed analysis by \citet{Huang-etal-21} is the observer-dependent
nature of their analysis.  We reiterate below, one more time, the
issue with the Eulerian, streamline-based analysis carried out by
\citet{Huang-etal-21}, no matter how sophisticated it is made (e.g.,
through lagged correlations and the construction of Hovm\"oller
plots of stitched fields on one side and of the other of the Lesser
Antilles) or how arbitrary it is framed (e.g., by following a
particular isoline of SSH).

\begin{figure}[t!]
  \centering%
  \includegraphics[width=.65\textwidth]{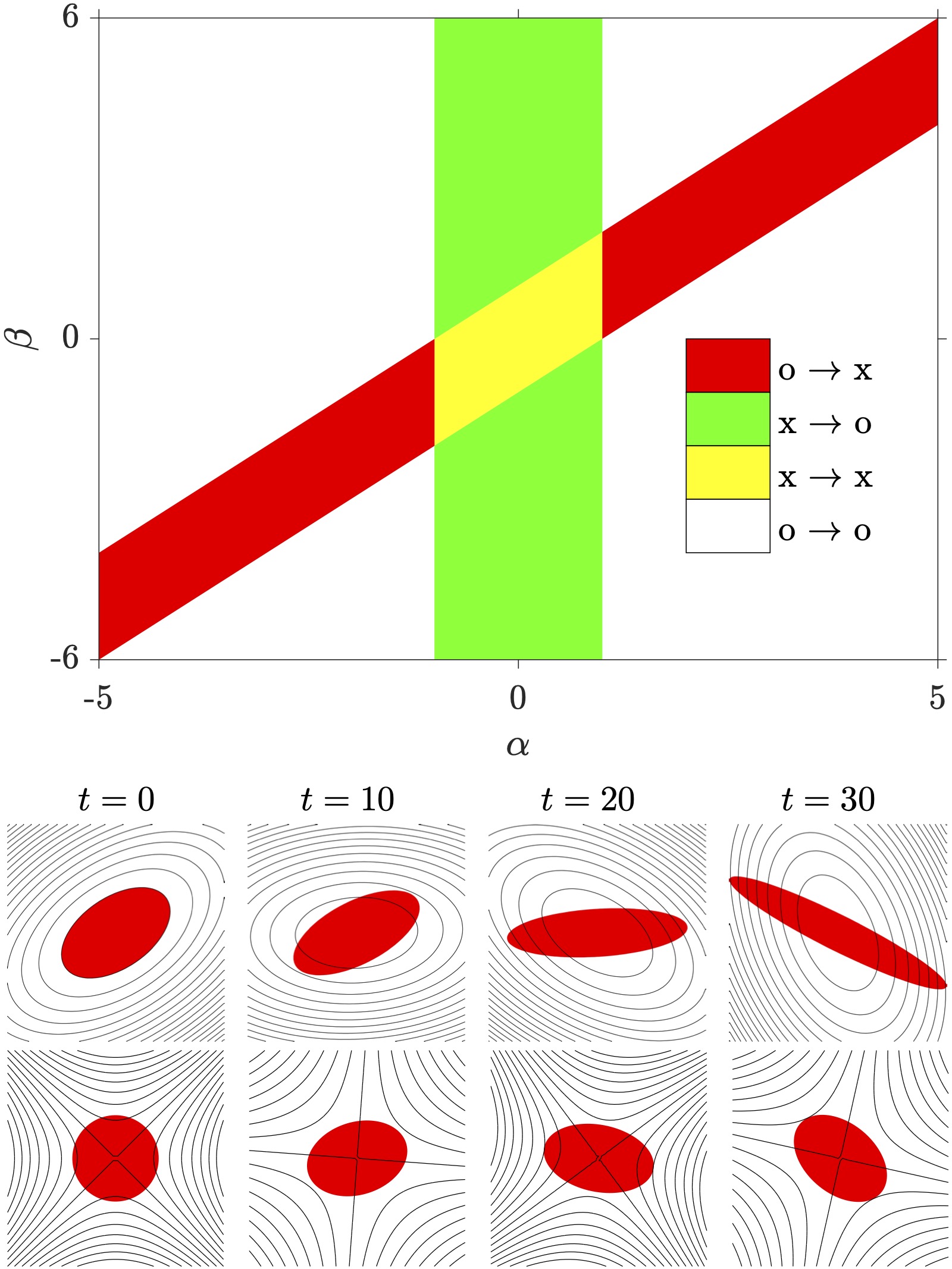}%
  \caption{(top panel) For the velocity field $\mathbf u(\mathbf
  x,t) = (x\sin2\beta t + y(\alpha+\cos2\beta t), x(\cos2\beta
  t-\alpha) - y\sin2\beta t)$, exact solution of the Navier-Stokes
  equation in two-space dimensions, complete possible streamline
  topology transformations under the observer change  $\mathbf x
  \mapsto \bar{\mathbf x}(\mathbf x,t) = (x \cos\beta t - y\sin\beta
  t, x\sin\beta t + y \cos\beta t)$, where ``o'' refers to center
  and ``x'' to saddle. (middle panels) Snapshots the evolution of
  a passive tracer (red) initially within a closed streamline of
  the velocity field with $(\alpha,\beta) = (2,2)$.  (bottom panels)
  As in the middle panels, but with $(\alpha,\beta) = (0,4)$ }
  \label{fig:vortex}%
\end{figure}

Consider the velocity field $\mathbf u(\mathbf x,t) = (x\sin4t +
y(2+\cos4t), x(\cos4t-2) - y\sin4t)$, where $\mathbf x = (x,y)\in
\mathbb R^2$ denotes position and $t\in\mathbb R$ is time, which
represents an exact linear solution of the two-dimensional
Navier--Stokes equation \citep{Haller-05}.  The flow streamlines are
closed at all times suggesting an elliptic structure, i.e., a vortex.
Moreover, the most common Eulerian criteria ($Q$, $\lambda_2$,
$\Delta$, swirling strength, etc., cf.\ \citet{Haller-05}) invariably
suggest that the flow holds a vortex.  In particular, the de-facto
oceanographic eddy detection diagnostic \citep{Chelton-etal-11a},
following a simple-minded argument by \citet{Flierl-81}, classifies
the flow as a ``nonlinear vortex'' since $U/c = \infty$, where $U$
is the rotational speed around the vortex and $c$ is the (here
vanishing) propagation speed of the vortex. However, if a tracer
is initialized within one instantaneously closed streamline, the
tracer does not remain trapped within the streamline as time goes
by, but rather it stretches bypassing all instantaneously closed
streamlines (Fig.\ \ref{fig:vortex}, middle panels).

The behavior just described has an explanation: this flow actually
hides a rotating saddle (pure deformation), as it follows by making
$\mathbf x \mapsto \bar{\mathbf x}(\mathbf x,t) = (x\cos2t - y\sin2t,
x\sin2t + y\cos2t)$, under which $\mathbf u(\mathbf x,t) \mapsto
(\bar y,\bar x) \equiv \bar{\mathbf u} (\bar{\mathbf x})$. The
$\bar{\mathbf x}$-frame is special inasmuch the flow in this frame
is steady, and thus flow streamlines and fluid trajectories coincide.
Hence short-term exposition pictures of the velocity field by the
observer in the $\bar{\mathbf x}$-frame determine the long-term
fate of fluid particles.  The only additional observation to have
in mind to fully determine the Lagrangian motion is that the observer
in the $\bar{\mathbf x}$-frame rotates (at angular speed $2$).  This
tells us that the flow under consideration is not actually unsteady
as there is a frame (the $\bar{\mathbf x}$-frame) in which it is
steady.

The situation can be made even more dramatic if the more general
linear solution to the two-dimensional Navier--Stokes equation is
considered: $\mathbf u(\mathbf x,t) = (x\sin2\beta t + y(\alpha+\cos2\beta
t), x(\cos2\beta t-\alpha) - y\sin2\beta t)$ where $\alpha,\beta
\in \mathbb R$ are arbitrary constants.  For instance, for
$(\alpha,\beta) = (0,4)$ one observes in the $\beta$-spinning
$\mathbf x$-frame a fixed saddle, while when viewed in the rotating
frame $\mathbf x \mapsto \bar{\mathbf x}(\mathbf x,t) = (x \cos\beta
t - y\sin\beta t, x\sin\beta t + y \cos\beta t) = (x \cos 8 t -
y\sin 8 t, \bar x\sin 8 t + \bar y \cos 8 t)$, one observes a center,
meaning that a seemingly purely deforming flow actually is a pure
rotation flow structure! That is, a material vortex. This is
illustrated in the bottom panels of Fig.\ \ref{fig:vortex}.  The
complete possible streamline topology transformations in parameter
space are presented in top panel of Fig.\ \ref{fig:vortex}, where
``o'' refers to center  (pure rotation) and ``x'' to saddle (pure
deformation).  If the exact solution so far discussed is considered
``pathological'' for being linear, \citet{Pedergnana-etal-20} discuss
a \emph{nonlinear} solution which exhibits behavior similar to that
just described.

In a truly unsteady flow there is no such distinguished observer
for whom the flow is steady \citep{Lugt-79}.  Thus one can never be
sure which observer gives the right answer when Eulerian vortex
detection---particularly the de-facto oceanographic vortex detection
diagnostic \citep{Chelton-etal-11a}---is applied.  As a consequence,
neither false positives nor false negatives can be ruled out
\citep{Beron-etal-13}, and thus life expectancy
estimates drawn from such diagnostics are unreliable \citep{Andrade-etal-20}.

In this Letter we show the results of applying a Lagrangian,
observer-independent eddy detection technique, known as \emph{geodesic
eddy detection} \citep{Haller-Beron-13, Haller-Beron-14}, on the
altimetry record over 1995-2019. This, of course, goes way beyond
observing closed streamlines of the altimetric SSH field
\citep{Chelton-etal-11a}. We reveal a number of NBCRs with material
boundaries capable of dragging material along.  However, no NBCR
is found to bypass the Lesser Antilles as a coherent material
structure.  The NBCRs have relatively short-lived coherent material
cores, experiencing strong breakaway filamentation before reaching
the Lesser Antilles, consistent with the drifter and float observations
by \citet{Fratantoni-Richardson-04}.

\section{Methodology}

\subsection{Geodesic Eddy Detection}

Let $\mathbf u(\mathbf x,t)$ be a two-dimensional fluid velocity,
with $\mathbf x$ denoting position in some domain of $\mathbb R^2$
and $t \in \mathbb R$ referring to time. Let $\mathbf F_{t_0}^t(\mathbf
x_0)$ denote the solution of fluid particle motion equation
\begin{equation}
  \dot{\mathbf x} = \mathbf u(\mathbf x,t)
\end{equation}
for initial condition $\mathbf x(t_0) = \mathbf x_0$.  Note that
$\mathbf F_{t_0}^t$ relates time-$t_0$ and $t$ positions of fluid
particles, and thus is called a \emph{$(t_0,t)$-flow map}.

The notion of a vortex with a material boundary resisting stretching
under advection by the flow, e.g., as inferred geostrophically from
altimetry data, from time $t_0$ to time $t_0 + T$ for some (finite)
$T$ is expressed by the variational principle \citep{Haller-Beron-13,
Haller-Beron-14}:
\begin{equation}
  S[\mathbf r] := \oint 
  \frac
  {\sqrt{\big\langle\mathbf r'(s), \mathsf C_{t_0}^{t_0+T}(\mathbf
  r(s))\mathbf r'(s)\big\rangle}}
  {\sqrt{\langle\mathbf r'(s), \mathbf r'(s)\rangle}}\,ds
  = 0,\quad 
  \left.\frac{d}{d\epsilon}\right\vert_{\epsilon = 0} S[\mathbf
  r + \epsilon \mathbf n] = 0.
  \label{eq:variational}
\end{equation}
Here, $\mathbf r(s)$ provides a parametrization for a material loop
at time $t_0$, $\mathbf n(s)$ is normal to the curve,
\begin{equation}
  \mathsf C_{t_0}^{t_0+T}(\mathbf x_0) := D\mathbf
  F_{t_0}^{t_0+T}(\mathbf
  x_0)^\top D\mathbf F_{t_0}^t(\mathbf x_0)
\end{equation}
is the (coefficient of the symmetric, positive-definite, right)
\emph{Cauchy--Green strain tensor field}, and the angle bracket
denotes scalar product. The integrand in \eqref{eq:variational},
which objectively (i.e., independent of the observer's viewpoint)
measures relative stretching from $t_0$ to $t_0 + T$, is invariant
under $s$-shifts and thus represents a Noether quantity.  That is,
it is equal to a positive constant, say $p$.  In other words,
solutions to \eqref{eq:variational} are characterized by uniformly
$p$-stretching loops. Embedded within $O(\varepsilon)$-thick coherent
material belts showing no observable variability in averaged relative
stretching, the time-$t_0$ positions of such \emph{$p$-loops} turn
out to be limit cycles of one of the following two bidirectional
vector or \emph{line} fields:
\begin{equation}
  \mathbf l_p^\pm(\mathbf r) :=
  \sqrt{
  \frac
  {\lambda_2(\mathbf r) - p^2}
  {\lambda_2(\mathbf r) - \lambda_1(\mathbf r)}
  }
  \,\mathbf{v}_1(\mathbf r) 
  \pm
  \sqrt{
  \frac
  {p^2 - \lambda_1(\mathbf r)}
  {\lambda_2(\mathbf r) - \lambda_1(\mathbf r)}
  }
  \,\mathbf{v}_2(\mathbf r),  
  \label{eq:p-line}
\end{equation}
where $\lambda_1 < p^2 < \lambda_2$.  Here, $\{\lambda_i\}$ and
$\{\mathbf v_i\}$, satisfying $0 < \lambda_1 \le \lambda_2$, $\langle
\mathbf v_i, \mathbf v_j\rangle = \delta_{ij}$, $i,j = 1,2$, are
eigenvalues and (orientationless) normalized eigenvectors, respectively,
of $\smash{\mathsf C_{t_0}^{t_0+T}}(\mathbf x_0)$.  Limit cycles
of \eqref{eq:p-line} either grow or shrink under changes in $p$,
forming smooth annular regions of nonintersecting loops.  The
outermost member of such a band of material loops is observed
physically as the boundary of a \emph{coherent Lagrangian eddy}.
The $p$-loops can also be interpreted as so-called null-\emph{geodesics}
of the (sign-indefinite) \emph{generalized Green--Lagrangian tensor
field}, $\mathsf C_{t_0}^{t_0+T}(\mathbf x_0)-p\,\mathsf{Id}$
\citep{Haller-Beron-14}.  For this and related results
\citep{Haller-Beron-12, Beron-etal-13, Haller-15} we refer to
this method of vortex identification as \emph{geodesic eddy detection}.

Material loops characterized by $p = 1$ resist the universally
observed material stretching in turbulence: they reassume their
initial perimeter at time $t_0+T$.  This conservation of perimeter,
along with the conservation of the enclosed area in the incompressible
case, conveys extraordinary coherence to geodesically detected
eddies.

An important observation is that the boundaries of coherent material
vortices revealed from geodesic detection are not only stretching
resisting, but also are nearly diffusion resisting \citep{Haller-etal-18}.
This may be expected for material fluid lines that do not experience
folding, and it is reassuring that it can be formally proved.

\subsection{Numerical Implementation}\label{sec:num}

Clearly, making reliable assessments of eddy material transport
requires one to pay a computational price. Geodesic eddy detection
is now well-documented \citep{Haller-Beron-13, Haller-Beron-14,
Karrasch-etal-14, Karrasch-Schilling-20} and software tools are
available.  We call the attention of the reader to the
\url{Julia} implementation employed here, \url{CoherentStructures.jl}. 
Nevertheless, we consider instructive to briefly summarize the algorithmic 
steps involved in geodesic eddy detection, enabling the practitioner to 
implement them in his/her preferred programming language:
\begin{enumerate}
  \item Provide $\mathbf u(\mathbf x,t)$ with $(\mathbf x,t) \in
	 (\mathcal D,\mathcal I) \subset \mathbb R^2 \times \mathbb
	 R$ and fix $t_0 \in \mathcal I$ and $|T| <
	 \text{length}\,(\mathcal I)$.  Here we use $\mathbf u =
	 gf^{-1}\hat{\mathbf z} \times \nabla \eta$, where
	 $g\hat{\mathbf z}$ is gravity, $f$ is the Coriolis parameter,
	 and $\eta(\mathbf x,t)$ is the altimetric SSH (i.e., with
	 mean dynamic topography added), as provided by the Archiving,
	 Validation and Interpretation of Satellite Oceanographic
	 data (AVISO+).
  \item Integrate $\mathbf u$ over $t\in [t_0,t_0+T]$ for
	 initial conditions $\mathbf x_0$ on fine lattice $\mathcal
	 G$ covering $\mathcal D$ well to get $\mathbf
	 F_{t_0}^{t_0+T}(\mathbf x_0)$ using some high-order method,
	 such as the Runge--Kutta family.
  \item Finite differentiate $\mathbf F(\mathbf x_0)$ over $\mathcal
	 G$ to get $\mathsf C_{t_0}^{t_0+T}(\mathbf x_0)$ and compute
	 $\mathsf C_{t_0}^{t_0+T} \xi = \lambda\xi$.
  \item Construct the $p$-line fields $\mathbf l_p^\pm$. 
  \item Using index theory for planar line fields
	 \citep{Karrasch-etal-14},  identify $\mathcal D_s \subset
	 \mathcal D$ with $\mathrm{ind}_{\partial\mathcal D_s}
	 (\mathbf l_p^\pm) = 1$, viz., with $\mathbf x_0^*$ satisfying
	 $\mathsf C_{t_0}^{t_0+T}(\mathbf x_0^*) = \mathsf{Id}$
	 such that the number of  wedges ($\mathrm{ind} = \frac{1}{2}$)
	 exceeds that of trisectors ($\mathrm{ind} = -\frac{1}{2}$)
	 by 2 (wedges and trisectors are the only type of singular
	 points, analogous to critical points of vector fields, in
	 two-space dimensions).  This is a necessary condition for
	 the existence of a limit cycle of the $p$-line field
	 \citep{Karrasch-etal-14}.
  \item Identify fixed points in Poincare sections transversal
	 to $\mathbf l_p^\pm$ in each $\mathcal D_s$ and keep the
	 outermost of all.  Each corresponding limit cycle defines
	 the boundary of a coherent material vortex that defies
	 stretching over $t\in [t_0,t_0+T]$.
\end{enumerate}

A refined version of this algorithm, proposed in \citet{Andrade-etal-20}
and utilized here, exhaustively searches the two-dimensional parameter
space $(t_0,T)$, \emph{allowing to frame genesis and apocalypse}.
More specifically, the refinement consists in repeatedly applying
geodesic eddy detection on the (altimetric) flow domain of definition
as follows:
\begin{enumerate}%[resume]
  \item[7.] Roll the initial time instance $t_0$ over a time window
  covering the time interval of during which a vortex is expected
  to exist.
  \item[8.] For each $t_0$, progress $T$ as long as a coherent Lagrangian
  eddy is successfully detected.  This way, for each $t_0$ a life
  expectancy $T_{\exp}(t_0)$ is obtained, which is the maximum $T$
  for which a geodesic eddy over $[t_0,t_0+T]$ is successfully
  detected. The expected result is a wedge-shaped $T_{\exp}(t_0)$
  distribution, indicating that all Lagrangian coherence assessments
  predict the breakdown consistently, independent of any parameter
  presets.  The \emph{birth date}, $t_\mathrm{bday}$, of the vortex
  is given by the $t_0$ for which $T_{\exp}(t_0)$ is maximized.
  The \emph{decease date} is then $t_\mathrm{bday} +
  T_{\exp}(t_\mathrm{bday})$.
\end{enumerate}
Robust assessments of the birth and decease dates of the vortex may
be obtained by combining the results from running the algorithm in
forward- and backward-time directions.

\section{Results}

An important cautionary remark is in order regarding using currents
derived geostrophically from altimetry in our analysis.  Depending
of how parameters are selected,  the validity of the (quasi)geostrophic
modeling framework may be brought as close to, or as far away from,
the equator as desired.  We here have chosen to stand middle ground
between being too permissive and too stringent by setting the lower
limit of the domain of analysis to 2$^{\circ}$N.  At this latitude
the Rossby number is about 0.7, as one can infer it based on
independent satellite drifter and float tracking estimates of NBCR
typical diameters (249 km) and azimuthal velocities (90 cm\,s$^{-1}$)
by \citet{Fratantoni-Richardson-04}.

\begin{figure}[t!]
  \centering%
  \includegraphics[width=\textwidth]{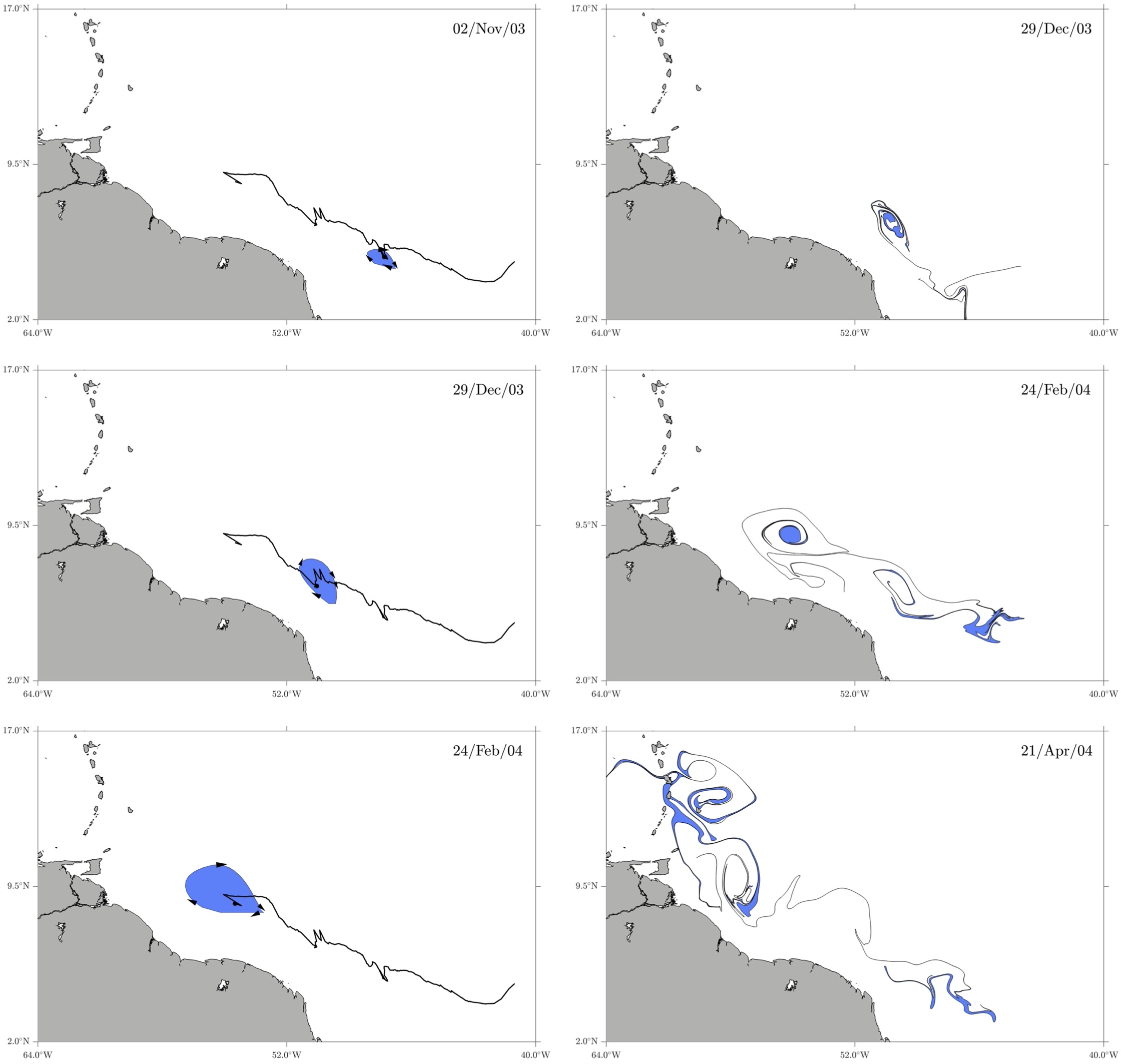} 
  \caption{(left
  panels) Selected snapshots of an NBCR classified as a nonlinear
  SSH eddy by the AVISO+ META database over 13/Aug/03 through
  24/Feb/04. The black curve in each panel corresponds to track
  \#74667 from this dataset. The arrows indicate the instantaneous
  direction of the flow along the SSH streamline that defines the
  boundary of the eddy on each day shown.  (right panels) Advected
  images after 57 days of each of the SSH eddy snapshots in the
  left panel under the altimetry-derived flow.}
  \label{fig:ssh}%
\end{figure}

Before applying geodesic eddy detection, we consider track \#74667
the AVISO+ Mesoscale
Eddy Trajectory Atlas Product (META) \citep{Pegliasco-etal-22}
to show that the tracked SSH feature neither bypasses the Lesser
Antilles as implied by \citet{Huang-etal-21}, nor it coherently
transports fluid. The former is evident from the naked-eye inspection
of track \#74667, which is shown in the left column of Fig.\
\ref{fig:ssh} along with 57-day-apart snapshots of the corresponding
SSH eddy, taken to be the region bounded by the outermost closed
SSH level set following \citet{Chelton-etal-11a}.  Track \#74667
never reaches the Lesser Antilles, and thus prolonging this trajectory
into the Caribbean Sea as in Figs.\ 2a--b of \citet{Huang-etal-21}
is indefensible.  Moreover, the instantaneous geostrophic velocity
field rotates anticyclonically along track \#74667, and the vector
field shown by \citet{Huang-etal-21} immediately west of the Lesser
Antilles inside the eastern Caribbean rotates cyclonically! On the
other hand, the nonlinear parameter for this SSH eddy $10 < U/c <
25$ approximately from 13/Aug/03 through 24/Feb/04.  However, it
does not hold within and transport along fluid coherently as
instructed by the de-facto oceanographic eddy detection method
\citep{Chelton-etal-11a}.  This is demonstrated in the right column
of Fig.\ \ref{fig:ssh}, which shows forward advected images after
57 days of each SSH eddy snapshot in the left column under the
altimetry-derived flow.  Note the vigorous filamentation experienced
by each SSH eddy snapshot upon advection.  Clearly, the fluid mass
initially inside the SSH eddy snapshot on 13/Aug/03 is not mapped
into that of 29/Dec/03 and neither is the mass initially the latter
into that on 24/Feb/04.  The flow is (nearly) incompressible, so
from the change of size of the SSH feature tracked is already quite
obvious, without the need of advection whatsoever, that one is not
in the presence of a coherent material vortex.  However, there is
evidence of material being trapped inside the ``core'' of the SSH
eddy while tracked over 29/Dec/03--24/Feb/04.  This suggests, and
we confirm below, that its positions during this time period are
intersecting those of a coherent Lagrangian vortex.  The SSH eddy
snapshot on 24/Feb/04 was forward-advected 57 days despite track
\#74667 ends on 24/Feb/04 to check if fluid initially inside traverses
the Lesser Antilles. Some fluid is indeed seen to penetrate into
the Caribbean Sea, consistent with direct measurements of biogeochemical
tracers \citep{Johns-etal-14}.  But this is accompanied by substantial
filamentation, even southeastward, more consistent with the observed
resistance of drifters and floats to bypass the Lesser Antilles
\citet{Fratantoni-Richardson-04} than with NBCRs connecting the
tropical Atlantic with the Caribbean Sea \citep{Huang-etal-21}.

\begin{figure}[t!]
  \centering%
  \includegraphics[width=\textwidth]{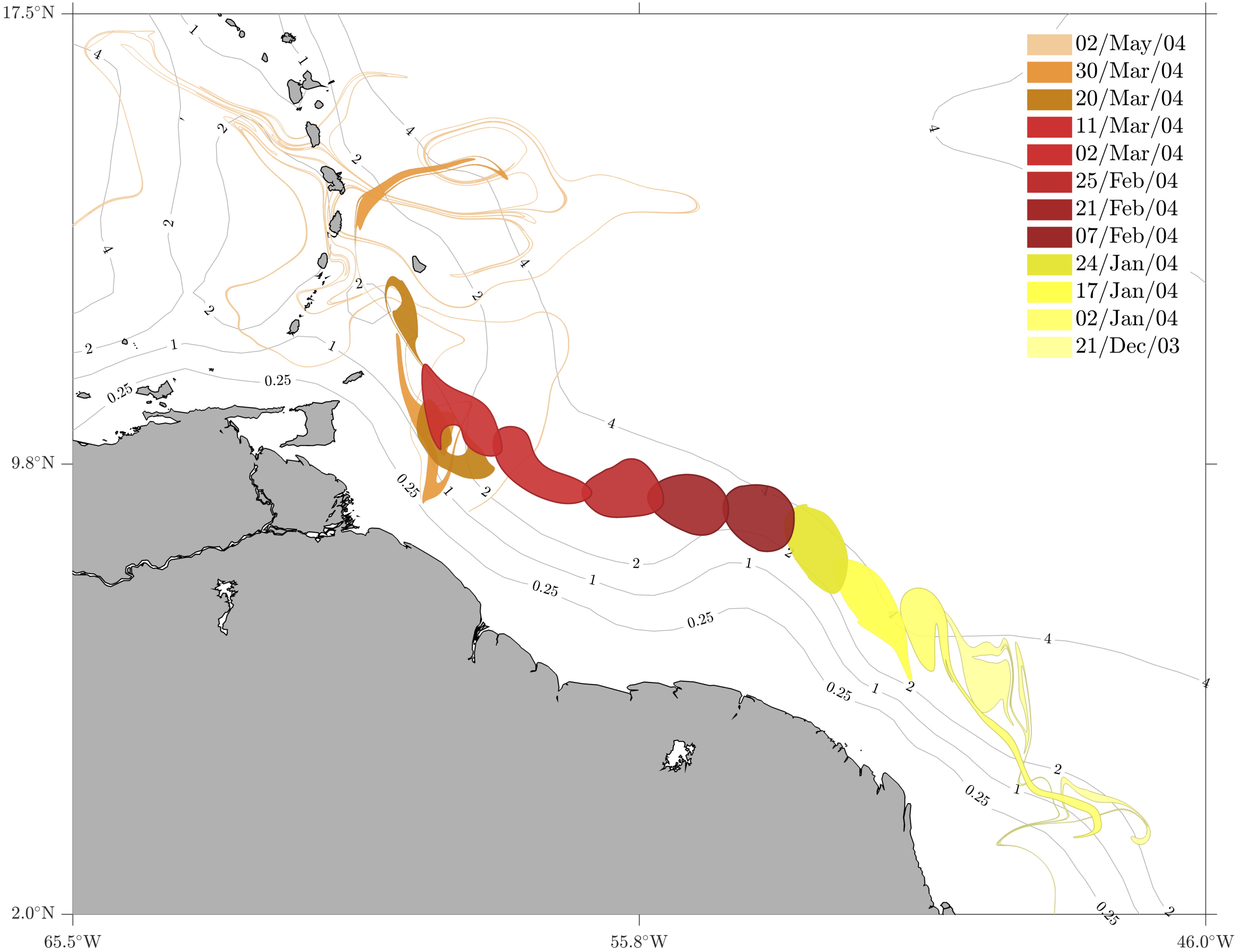}%
  \caption{Genesis (yellow), evolution (red), and apocalypse (orange)
  of a coherent Lagrangian NBCR, geodesically detected from
  altimetry-derived velocity data using the SSH eddy
  trajectory of Fig.\ \ref{fig:ssh} as a reference.  Selected
  isobaths (in km) are shown in gray.}
  \label{fig:gen}%
\end{figure}

We now proceed to applying geodesic eddy detection. Starting by
taking SSH eddy track \#74667 as a reference, we were able to extract
an anticyclonic coherent Lagrangian vortex, identifiable with an
NBCR. The vortex acquired material coherence on $t_\mathrm{bday}
=$ 07/Feb/04, which lasted until 20/Mar/04.  This corresponds to a
rigorous material coherence life expectancy of $T_{\exp}(t_\mathrm{bday})
= 42$ days, which maximizes the function $T_{\exp}(t_0)$ obtained
by rolling the coherence assessment time, $t_0$, over the duration
of SSH eddy track \#74667, and progressing the integration time,
$T$, until no vortex boundary is geodesically extracted.  This
explains why some material was seen to be trapped inside the SSH
eddy corresponding to track \#74667: the SSH eddy snapshot on
29/Dec/03 intersects the advected image on that date of the
geodesically detected vortex on $t_\mathrm{bday} =$ 07/Feb/04. This
vortex falls on the shorter side of the mesoscale spectrum, with
an \emph{equivalent radius} $r_\mathrm{eq} = 35$ km.  We define
$r_\mathrm{eq}$ as the average over the time interval
$[t_\mathrm{bday},t_\mathrm{bday} + T_{\exp}(t_\mathrm{bday})]$ of
the radii of the circumferences whose perimeters are equal to those
of the forward-advected images over that interval of the boundary
of the extracted coherent Lagrangian vortex on $t_\mathrm{bday}$.
The perimeter of the material boundary of the NBCR in question is
further characterized by growing over its lifespan by a factor $p
= 2$.  This exemplifies well its characteristic property, namely,
a strong tendency to resist stretching.  In Fig.\ \ref{fig:gen} we
illustrate the life cycle of the geodesically detected NBCR,
indicating in red tones the material vortex while is classified as
coherent.  In yellow tones we show several backward-time images of
the NBCR on $t_\mathrm{bday} =$ 07/Feb/04, illustrating its genesis.
This happens at around 52$^{\circ}$W, consistent with assessments
of NBCR formation from the analysis of satellite-derived chlorophyll
images \citep{Fratantoni-Glickson-02} and also of float and drifter
trajectories \citep{Fratantoni-Richardson-04}. How material coherence
is acquired is a formidable problem, which admittedly lies beyond
the reach of the nonlinear dynamics tool employed here.  We note,
however, that it must build on some underlying physical process.
Priorly proposed ``NBCR pinch off'' mechanisms include potential
vorticity transformation in the nonlinear cross-equatorial flow of
the NBC \citep{Edwards-Pedlosky-98, Zharkov-Nof-10} and barotropic
instability of the North Equatorial Counter Current and ensuing
radiation of Rossby waves \citep{Jochum-Malanotte-03}.  Several
forward-advected images of the NBCR on $t_\mathrm{bday} +
T_{\exp}(t_\mathrm{bday}) =$ 20/Mar/04 are depicted in orange tones.
This illustrates the apocalypse of the NBCR, whose contents do not
traverse the Lesser Antilles coherently.  Indeed, the NBCR loses
material coherence several hundreds of km away from the Antilles
Arc, when encountering water depths shallower than 2000 m, which
applies to many other geodesically detected NBCRs as we show below.
According to in-situ hydrography \citep{Fratantoni-etal-95}, the
anticyclonic circulation of NBCRs may be traced down in the water
column as deep as 1000 m.  This seems sufficiently deep for them
to become unstable due to topographic effects.  But surface-intensified
NBCRs have also been observed \citep{Wilson-etal-02}, suggesting
that they may bypass topography without difficulty \citep{Adams-Flierl-10}
and thus other mechanisms, such as thermal instability as proposed
by \citet{Andrade-etal-22} for Caribbean Sea eddies, might be
contributing to their loss of material coherence.  This is a much
more tractable problem than material coherence acquisition, and is
reserved for near future investigation. (Subsurface NBCRs have also
been identified \citep{Johns-etal-03}, but these are not detectable
from altimetry and thus are beyond the scope of this paper.)

\begin{figure}[t!]
  \centering%
  \includegraphics[width=\textwidth]{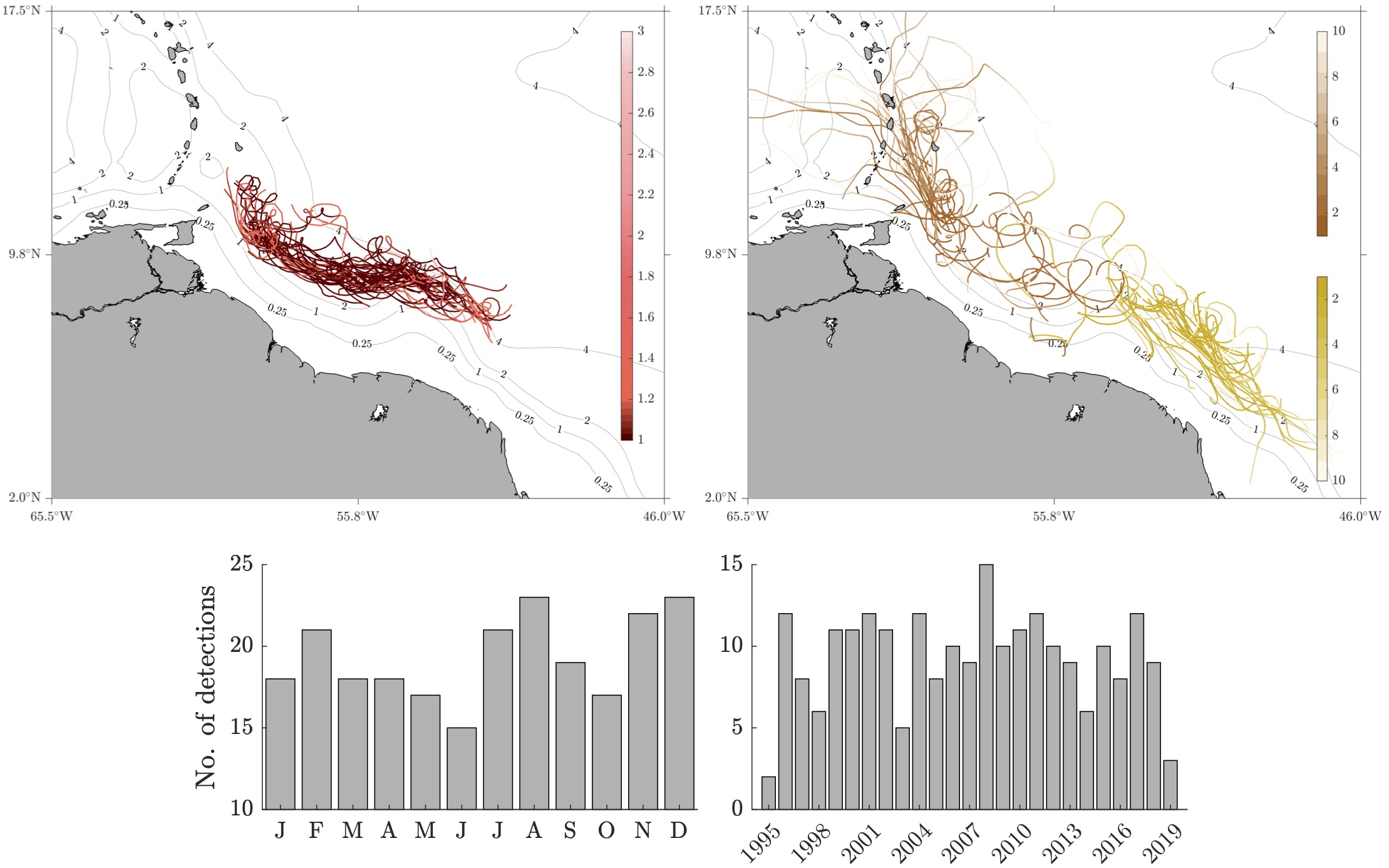}%
  \caption{(top-left panel) Trajectories of the centroids of
  geodesically detected NBCRs over 1995--2019 from altimetry-derived
  velocity with colors indicating stretching factor ($p$).  (top-right
  panel)  Thirty-day-backward (resp., forward) trajectories of the
  centroids of the regions occupied by the geodesically detected
  NBCRs on their birth (resp., decease) date.  The backward (resp.,
  forward) trajectories are depicted using yellowish (resp., brownish)
  colors, whose tonalities decrease with increasing relative
  stretching of the boundary of the advected fluid regions.
  (bottom-left) Number of coherent Lagrangian NBCRs detected as a
  function of the month. (bottom-right) As in the left, but as a
  function of the year.}
  \label{fig:cen}%
\end{figure}

Geodesic eddy detection is finally applied over a long period of
time, 1995 through 2019.  A census of coherent Lagrangian NBCRs is
constructed, independent of reference SSH eddy tracks.  More
specifically, we repeatedly search for coherent Lagrangian NBCRs
by applying geodesic eddy detection inside a $7^\circ \times 8^\circ$
box centered at (9$^\circ$N,54$^\circ$W).  While this is a region
where NBCR shedding can be expected to take place
\citep{Fratantoni-Glickson-02, Fratantoni-Richardson-04}, if a
vortex is geodesically detected on some $t_0$ for some \emph{trial}
integration time $T$, its centroid is advected in \emph{backward}
time until geodesic detection applied in \emph{forward} time does
not detect any vortex, enabling framing genesis with precision.
Steps 7 and 8 of the geodesic eddy detection algorithm (Sec.\
\ref{sec:num}) are then applied starting on the last $t_0$ for which
a vortex is detected.  Rolling $t_0$ forward from that $t_0$ does
not require any SSH eddy trajectory as a reference, as the trajectory
of the centroid of the geodesically detected vortex is employed.
In the top-left panel of Fig.\ \ref{fig:cen} we show the trajectories
of the centroids of all geodesically detected NBCRs. These are
colored according to their stretching factor ($p$).  Out of a total
of 196 NBCRs detected over 1995--2019, none was seen to reach the
Lesser Antilles as a coherent material vortex.  Rigorous material
coherence is lost in many cases when a ring approaches the 1500-m
isobath or so. On average, coherent material NBCRs possess a life
time of $T_{\exp}(\mathrm{bday}) = 30 \pm 18$ d, stretch by a factor
of $p = 1.2 \pm 0.35$, and have an equivalent radius of $r_\mathrm{eq}
= 42 \pm 16$ km.  The bottom-left (resp., right) panel at the bottom
of Fig.\ \ref{fig:cen} shows the number of geodesically detected
NBCRs as a function of the month (resp., year).  While the number
varies (from 2 to 15) along the year, no statistically significant
seasonal (annual or annual-plus-semiannual) signal is seen to fit
the observed variability.  In turn, $9\pm 3$ coherent material NBCRs
are formed yearly on average, revealing no statistically significant
trend of any sign.  Finally, in the top-right panel of Fig.\
\ref{fig:cen} we show the trajectories of the centroids of the
regions occupied by the backward (resp., forward) advected images
of the initial (resp., final) positions of the boundaries of the
geodesically detected NBCRs for 30 days.  We use yellowish (resp.,
brownish) colors for the backward (resp., forward) trajectory pieces.
The tonality of these colors decreases with increasing relative
stretching of the advected boundaries, computed as the ratio of
initial-to-final advected boundary arclength.  Note, in particular,
that relative stretching increases considerably near the Lesser
Antilles, indicating that the contents of material NBCRs, once
coherent, get mixed with the ambient fluid in the Atlantic before
a typically small fraction thereof finds its way into the Caribbean
Sea.

We have intentionally avoided entering the debate over the reliability
of transport calculations based on altimetry-derived velocities,
with diverging views, cf., e.g., \citet{Poje-etal-17} as an example
of a critical view and \citet{Beron-LaCasce-16} with a contrasting
opinion. However, in response to the request of an anonymous reviewer,
we have included in the Supporting Information a figure (S1) which
provides support for the latter, more positive view, with a comparison
of a geodesic eddy detection based on velocities as provided by
AVISO+ on the standard 0.25$^\circ$ $\times$ 0.25$^\circ$ grid and
on velocities from the experimental Multiscale Interpolation Ocean
Science Topography (MIOST), which are higher-resolution velocities
on a 0.1$^\circ$ $\times$ 0.1$^\circ$ grid that combine AVISO+
velocities and velocities from available surface buoys from the
NOAA Global Drifter Program.  The details of the life cycle of the
NBCR geodesically detected using AVISO+ velocity are visually very
similar to those of the NBCR detected using MIOST velocity.
But there is theoretical reason to expect this.  Indeed,
while the flow-invariant boundaries of geodesic eddies are of
elliptic nature, they follow as limit cycles of a dissipative vector
(line) field, which are structurally stable \citep{Arnold-73}.
 
\section{Conclusions}

North Brazil Current Rings (NBCRs) do not connect the connect the
tropical Atlantic Ocean and the Gulf of Mexico as coherent material
vortices despite recent indefensible claims \citep{Huang-etal-21}.
Supported on direct quasi-Lagrangian observations by
\citet{Fratantoni-Richardson-04}, we have provided proof for our
rebuttal using \emph{geodesic eddy detection}, a Lagrangian method
rooted in geometric nonlinear dynamics \citep{Haller-Beron-13,
Haller-Beron-14}.  Geodesic vortices detected from satellite altimetry
have material (i.e., flow invariant) boundaries that defy stretching
in an otherwise typically turbulent ambient fluid. They remain the
same independent of whether the observer changes its stand point.
Moreover, these boundaries are minimizers of diffusion and stochastic
transport transversal to them \citep{Haller-etal-18}.  The main issue
with streamline-based, Eulerian eddy detection \citep{Chelton-etal-11a},
which forms the basis of the analysis of \citet{Huang-etal-21},
lies in its observer-dependent nature, which makes it useless in
drawing long-term conclusions about material transport.   This issue
has been brought up in oceanography many times with unfortunately
small reception since at least the work of \citet{Beron-etal-08-GRL}.

That some material dragged along by NBCRs that we have been able
to classify as material coherent for typically short periods of
time can bypass the Lesser Antilles is out of the question.
However, once again, this does not happen in an organized,
flow-invariant manner.  Very intriguingly, that material coherence
can be rebuilt in the eastern Caribbean Sea is possible and indeed
happens, as it has been shown for the first time recently by
\citet{Andrade-etal-22}, in connection with transportation of
\emph{Sargassum} and subsequent inundation of coastal areas.  The
squeezing-through-gaps and subsequent material buildup mechanism
proposed by \citet{Simmons-Nof-02} cannot be used to shed light on
the problem since NBCRs do not reach the Lesser Antilles as coherent
Lagrangian vortices.  Why and how material coherence out of incoherent
fluid manifests is a fascinating problem that, it seems, will take
many years to be elucidated.  A seemingly more tractable task is
assessing how much fluid initially carried by NBCRs is transported
into the Caribbean Sea, which is of interest for varied reasons.
Specialized tools to carry out such a task are available
\citep{Miron-etal-21-Chaos, Drouin-etal-22, Miron-etal-22}, which
will be considered in the future.

The main take-away message from this study is that the oceanographic
community, for the reasons exposed above, should exert care when
interpreting Eulerian mesoscale eddy data such as those distributed
by AVISO+ through the META database.

We close by responding to criticism by two anonymous reviewers,
referred to herein as R1 and R2. R1 argued that potential vorticity
(PV), as considered in \citet{Zhang-etal-14}, should be used in
coherent eddy detection instead of geodesic eddy detection as in
this paper.  R1 advocates the use PV for it being a ``dynamical''
quantity.  We do not ignore the importance of PV for the insight
into geophysical fluid dynamics it can provide.  However, PV cannot
be used to unequivocally frame material coherence because observers
that unsteadily rotate and translate relative to one another on the
planet will arrive at different conclusions, that is, PV is not
objective. This is because the total vorticity itself changes with
terrestrial observer's changes, as it is shown in the online
Supporting Information for the case of shallow-water PV on the
$\beta$-plane.  Furthermore, the use of PV in material coherence
assessment depends on the so-called PV-barrier argument
\citep{Dritschel-McIntyre-08}, which is valid for zonally-oriented
PV steps and small-amplitude perturbations to them.  Obviously, the
latter requires material conservation of PV, which in general cannot
be expected.  By contrast, geodesic eddy detection is objective and
is not constrained by PV conservation or any assumption beyond
two-dimensionality.  R2, in turn, suggested that the Navier--Stokes
solutions discussed in the Introduction are irrelevant for geophysical
flows.  The ``o-to-x'' behavior illustrated in the middle panels
of Fig.\@~1 is well-documented.  Indeed, this paper is one example
among many others \citep{Beron-etal-13, Haller-Beron-13,
Andrade-etal-20, Denes-etal-22}. The ``x-to-o''
behavior illustrated in the bottom panels of Fig.~1 is less common;
Fig.\@~S2 in the Supporting Information provides some evidence of
it.

\section*{Acknowledgements} 

The criticism of two anonymous reviewers has helped us driving the
message of our paper more clearly.  We thank Daniel Karrasch for
the benefit of many useful discussions on geodesic eddy detection
implementation and also George Haller for discussions on objectivity.
Technical support by A.\ Dominguez-Guadarrama is acknowledged.  FAC
was supported by ECOSUR core funding and FJBV by NSF grant OCE2148499.

\section*{Conflict of Interest}

The authors declare no conflicts of interest relevant to this study.

\section*{Open Research}

The gridded multimission altimeter products
\href{https://www.aviso.altimetry.fr/en/data/data-access.html}{https://\allowbreak
www.aviso.\allowbreak altimetry.fr/\allowbreak en/data/\allowbreak
data-access.html} including the experimental MIOST product
(\href{https://data.aviso.altimetry.fr/aviso-gateway/data/SLA_MIOST_alti_drifters/}{https://\allowbreak
data.aviso.\allowbreak altimetry.fr/\allowbreak aviso-gateway/\allowbreak
data/SLA\_MIOST\_alti\_drifters/}) were produced by SSALTO/DUACS
and distributed by AVISO+, with support from CNES.  The META database
(\href{https://data.aviso.altimetry.fr/aviso-gateway/data/META3.1exp_DT/}{https://\allowbreak
data.aviso.\allowbreak altimetry.fr/\allowbreak aviso-gateway/\allowbreak
data/\allowbreak META3.1exp\_DT/}) was produced by SSALTO/DUACS and
distributed by AVISO+, with support from CNES, in collaboration
with Oregon State University with support from NASA.

\appendix

\section{AVISO+ vs MIOST}

Figure S1 provides a comparison of the life cycle of an NBCR extracted
using geodesically detected from AVISO+ velocity data with that of
the same NBCR but as geodesically extracted from MIOST velocity
data.

\begin{figure}[t!]
  \centering%
  \includegraphics[width=.75\textwidth]{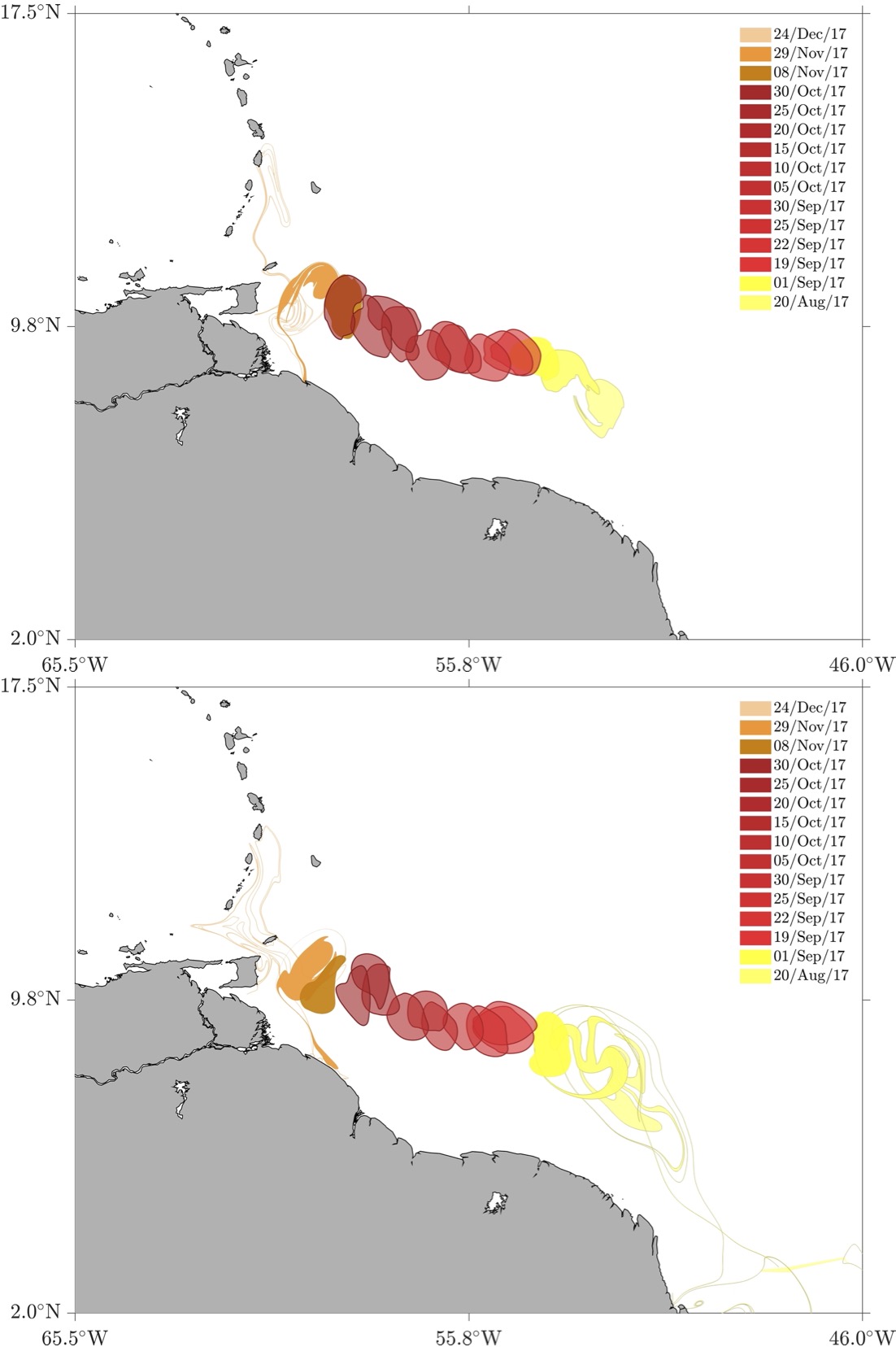}%
  \caption{Genesis (yellow), evolution (red), and apocalypse (orange)
  of a coherent Lagrangian NBCR, geodesically detected from
  AVISO+ (top) and MIOST (bottom) velocity data.}
  \label{fig:gen}%
\end{figure}

\section{Evidence of ``x-to-o'' behavior}

Figure S2 provides support to the ``x-to-o'' behavior illustrated
in the bottom panels of Fig\@~1 based on a geodesic eddy detection
from altimetry-derived velocity in the Gulf of Mexico.

\begin{figure}[t!]
  \centering%
  \includegraphics[width=.75\textwidth]{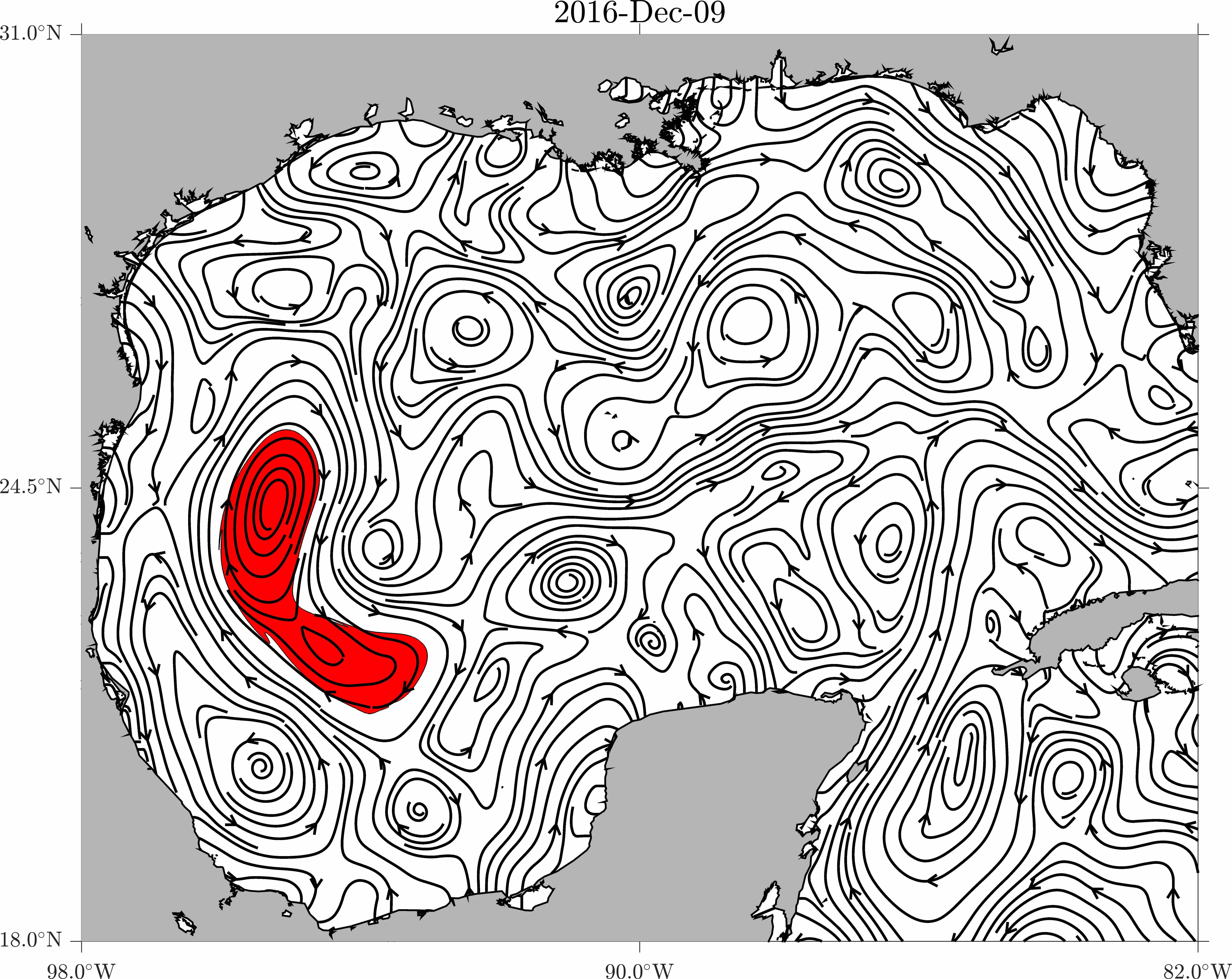}%
  \caption{Mesoscale geodesic eddy (red) detected from altimetry-derived
  velocity in the Gulf of Mexico with instantaneous streamlines of
  the field overlaid (black curves). Note the instantaneous
  ``x-point'' in the center of the coherent material eddy.}
  \label{fig:gen}%
\end{figure}

\section{Observer dependence of the potential
vorticity}

A scalar field $s(\x,t)$ on $\R^2 \times \R$ under the
change of observer
\begin{equation}
  \x \mapsto \bar\x =  
  \begin{pmatrix}
  \cos\alpha(t)  & \sin\alpha(t)\\ -\sin\alpha(t) & \cos\alpha(t)
  \end{pmatrix} \x + \b(t)
  \label{eq:T}
\end{equation}
transforms as \citep{Truesdell-Noll-65}
\begin{equation}
  \bar s(\bar\x,t) = s(\x,t).
\end{equation}
Now, on the $\beta$-plane, the total spin,
\begin{equation}
  \Omega_f = \tfrac{1}{2}\big(\nabla\u_f - (\nabla\u_f)^\top\big) =
  - \tfrac{1}{2}(\omega + f)\J,
  \label{eq:Of}
\end{equation}
where $\omega(\x,t) = \z\cdot\nabla\times\u(\x,t)$ is the relative
vorticity, $f(\x) = f_0 + \beta y$ is the Coriolis parameter, and
$\J$ represents an $\frac{1}{2}\pi$-anticlockwise rotation.  This
is satisfied by $\u_f(\x,t) = \u(\x,t) + \f(\x)$ for any $\f$ such
that $\z\cdot\nabla\times\f = f$, which represents the vector
potential of the local angular velocity of the planet, $\frac{1}{2}f\z$
\citep{Salmon-83}.  Under the transformation \eqref{eq:T} \emph{applied
on the $\beta$-plane}, \eqref{eq:Of} becomes
\begin{equation}
  \bar\Omega_f = -\tfrac{1}{2}\big(\omega - 2\dot\alpha + \varphi\big)\J
  \label{eq:barOf}
\end{equation}
for some $\varphi(\x,t)$ such that $\bar\nabla\bar\f -
(\bar\nabla\bar\f)^\top = - 2\varphi J$.  Equation \eqref{eq:barOf}
shows how the total vorticity, $(\omega + f)\z$, changes under a
change of $\beta$-plane observer \eqref{eq:T}. Note, in particular,
that $\dot\alpha$ is the angular velocity of the observer change.
The shallow-water Ertel's potential vorticity,
\begin{equation}
  q = \frac{\omega + f}{h},
  \label{eq:q}
\end{equation}
where $h(\x,t)$ is the fluid thickness, then transforms, under
\eqref{eq:T}, as
\begin{equation}
  \bar q = \frac{\omega - 2\dot\alpha + \varphi}{h},
\end{equation}
since $\bar h = h$, as this does not depend on velocity. Thus
\begin{equation}
  \bar q \neq q.
\end{equation}
This means that $q$ changes for terrestrial observers that unsteadily
rotate and translate relative to one another, and hence is not
objective.

\bibliographystyle{mybst-square}
\bibliography{fot}

\end{document}